\documentclass{PoS}

\title{The Mu2e Experiment at Fermilab}

\ShortTitle{The Mu2e Experiment at Fermilab}

\author{\speaker{Luca Morescalchi}\thanks{On the behalf of the Mu2e Collaboration, see \cite{MU2ECOLLABORATION}.}\\%
        University of Siena and INFN Pisa \\
        E-mail: \email{luca.morescalchi@pi.infn.it}}

\abstract{The Mu2e Experiment at Fermilab will search for the coherent, neutrinoless conversion of muons into electrons in the field of an aluminium nucleus with an unprecedented 
sensitivity. Such a charged lepton flavor-violating reaction probes new physics at a scale inaccessible with direct searches 
at either present or planned high energy colliders. Moreover, the Mu2e experiment both complements and extends the current search for the $\mu \to e \gamma$ decay at MEG 
and searches for new physics at the LHC. We will present the physics motivation for Mu2e, the experimental setup and the current status of the experiment.}

\FullConference{XXIV International Workshop on Deep-Inelastic Scattering and Related Subjects\\
		11-15 April, 2016\\
		DESY Hamburg, Germany}

\begin{document}

\section{Introduction}

\noindent Within the Standard Model (SM) heavy leptons decay in a way that almost perfectly conserves lepton family number. Even after the discovery of neutrino oscillations, in 
the minimal extension of SM, the predicted branching ratios of the Charged Lepton Flavor Violation (CLFV) processes in the muon sector are smaller than 10$^{-50}$, unreachable by 
the current particle accelerators. Starting from the first search performed by Hinks and Pontecorvo in 1948 \cite{Hincks:1948vr}, despite repeated efforts, no CLFV process has been observed yet; 
so any experimental observation would be an unambiguous signal of physics beyond the Standard Model. A complete review that covers details of the CLFV experimental 
history, the theoretical implications and the technical iussues can be found in \cite{Bernstein:2013hba} and in \cite{CeiA:2014wea}. One of the most promising processes for probing CLFV is the coherent 
muon conversion in the field of a nucleus, $\mu A \rightarrow e A$. In this process the nucleus is left intact and, due to the recoil, the resulting electron has a monochromatic 
energy slightly below the muon rest mass. The Mu2e experiment, concurrently with the COMET experiment being built at J-PARC \cite{Cui:2009zz}, is designed to improve the current limit on the 
ratio $R_{\mu e}$ by 4 orders of magnitude over SINDRUM II \cite{Bertl:2006up}. The ratio $R_{\mu e}$ is defined as:

\vspace{1.9 mm}

\begin{equation}
R_{\mu e} = \frac{\Gamma(\mu^{-} + N(A,Z) \rightarrow e^{-} + N(A,Z))}{\Gamma(\mu^{-} + N(A,Z) \rightarrow \textrm{all muon captures})} 
\end{equation}

\vspace{4.2 mm}

\noindent where, in the Mu2e case, $N(A,Z)$ is an aluminum nucleus. Many New Physics scenarios, like many SUSYs, Leptoquarks, Heavy Neutrinos, GUTs, Extra Dimensions or Little Higgs, 
predict significantly enhanced values for $R_{\mu e}$, at a level accessible by the expected Mu2e sensitivity.   

\vspace{1.9 mm}

\begin{figure}[htbp]
\centering
\includegraphics[width=0.80\textwidth]{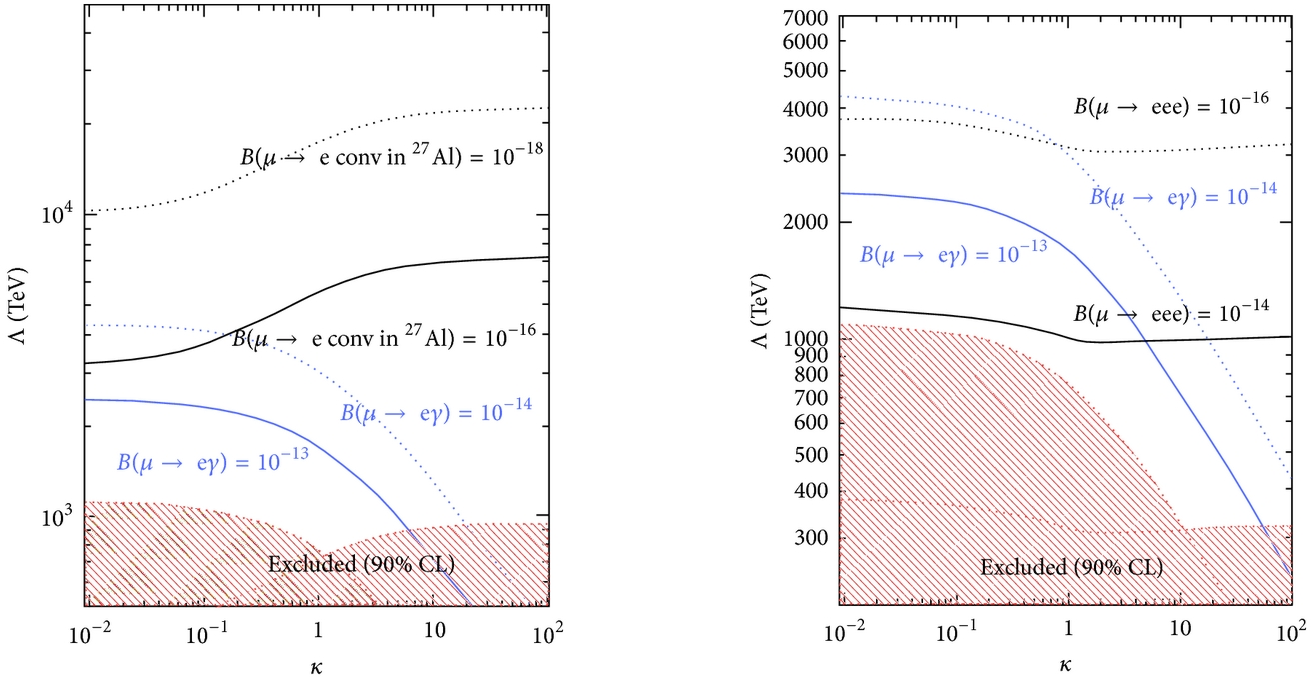}    
\caption{\scriptsize{\emph{Left:} Sensitivity of $\mu \rightarrow e\gamma$ (blue) and $\mu A \rightarrow e A$ conversion experiments (black) in the ($\kappa$, $\Lambda$) plane. 
\emph{Right:} Sensitivity of $\mu \rightarrow e\gamma$ (blue) and $\mu \rightarrow eee$ (black) in the same plane. The red regions are excluded by MEG for small $\kappa$ and by 
SINDRUM and SINDRUM II for large $\kappa$. \cite{CeiA:2014wea}} \label{img:lagrangian}}
\end{figure}

\vspace{1.9 mm}

\noindent To better understand the Mu2e discovery potential it is convenient to use a model-indipendent parametrization \cite{CeiA:2014wea} that can describe CLFV processes as an admixture of two classes 
of diagrams: the first one includes magnetic moment loop diagrams with a photon exchanged; the second one includes both contact terms and the exchange of a new heavy particle. The 
coefficients of these two types of operators are parameterized by two independent constants: $\Lambda$, the mass scale of the new physics, and $\kappa$, a dimensionless parameter that mediates 
between the two terms. Figure \ref{img:lagrangian} shows the exclusion plots at 90\% of C.L. for several CLFV processes involving the muon in the ($\kappa$,~$\Lambda$) plane. 
Mu2e explores mass scales up to 10$\ldotp$000~TeV, far beyond the scales that will be accessible to direct observation at the LHC. Moreover, Mu2e and $\mu \rightarrow eee$ experiments  
are sensitive over a wide $\kappa$ range, while the $\mu \rightarrow e \gamma$ experiments, like MEG, are not sensitive in case the contact term dominates.

\section{The Mu2e Experiment}

\begin{figure}[htbp]
\centering
\includegraphics[width=0.9\textwidth]{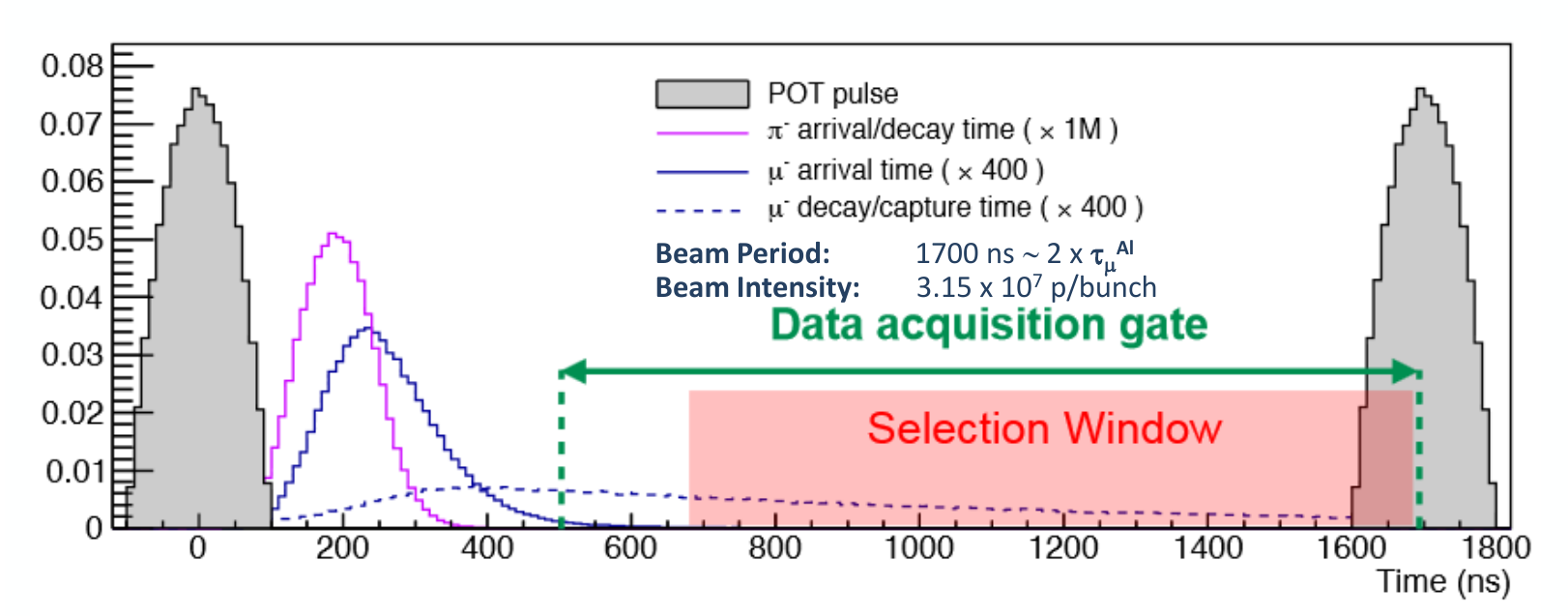}    
\caption{\scriptsize{The Mu2e spill cycle for the proton on target pulse and the delayed selection window that allows an effective elimination of the prompt background.} 
\label{img:beamstruct}}
\end{figure}

\noindent To reach the sensitivity goal $R_{\mu e}$ < 6 $\cdot$ 10$^{-17}$ @ 90$\%$ C.L., about 10$^{18}$ stopped muons are needed. Muons are produced using 8 GeV protons from the Fermilab 
accelerator complex, which provides a sequence of 200~nsec wide micro-bunches separated by 1.7~$\mu$sec. The beam period is roughly twice the muon mean lifetime in Al nucleus, $\tau_{\mu}^{Al}$ = 864~ns).
This particular beam structure, as shown in Figure \ref{img:beamstruct}, allows Mu2e to use a delayed selection windows to suppress the prompt background coming from proton 
interactions.    

\begin{figure}[htbp]
\centering
\includegraphics[width=1\textwidth]{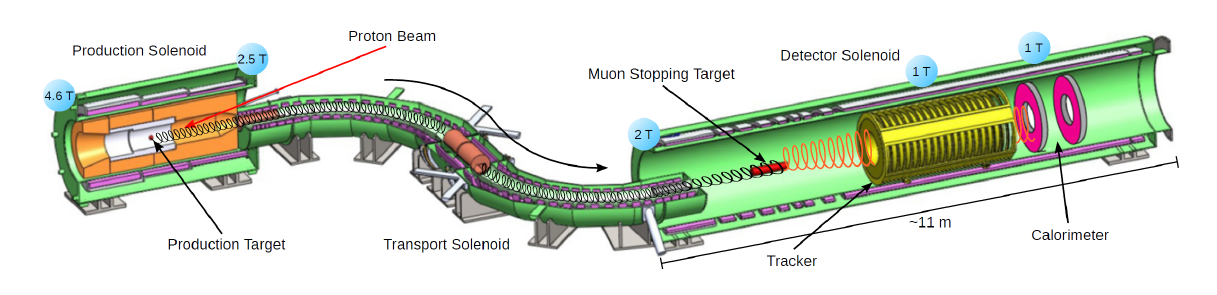}    
\caption{\scriptsize{Diagram of the Mu2e muon beam-line and detector. The cosmic ray veto system and the stopping target monitor are not shown in this figure.} \label{img:beamline}}
\end{figure}

\noindent The Mu2e apparatus consists of three superconductive solenoids, as shown in Figure \ref{img:beamline}: the Production Solenoid (PS), the 
Transport Solenoid (TS) and the Detector Solenoid (DS). The proton beam interacts in the PS with a tungsten target, producing mostly pions and muons.
A back-scattered muon beam is captured by the PS and transported through the S-bend TS, that selects low momentum negative charged particles and delivers them to the aluminum stopping 
targets in the DS. Electrons from the $\mu \to e$ conversion (CE) in the stopping target are captured 
by the magnetic field in the DS and transported through the Straw Tube Tracker, that reconstructs the CE trajectory and its momentum. The CE then strikes the Electromagnetic Calorimeter, 
that provides independent measurements of the energy, the impact time and the position. Both detectors operate in a 10$^{-4}$~Torr vacuum and in an uniform 1~T axial magnetic field. A 
cosmic ray veto system, which consists of four layers of scintillator bars, covers the entire DS and half of the TS, and guarantees a a veto inefficiency at the level of 10$^{-4}$.
Measurement of the total number of captured muons is provided by a high purity germanium detector, via the observation of the x-rays resulting from the nuclear capture. Additional 
details on the Mu2e apparatus can be found in \cite{MU2ETDR}.

\subsection{The Tracker}
\noindent Having the duty to identify the conversion electrons, the extremely low-mass tracker is the core of the experiment. It consists of about 20k straw tubes of 5 mm diameter filled 
with an Ar/CO$_{2}$ mixture. The tubes are arranged orthogonally to the solenoid axis and are grouped, over 3.2~m, 
in 18 tracking stations that have the inner 38 cm left un-instrumented. Only particles with a P$_{T}$>55~MeV reach the detector and only particles with P$_{T}$>90~MeV  
leave enough hits to form a reconstructible trajectory. In this way, the tracking system is blind to the majority of the background particles coming from the muons interactions in the 
stopping target. The cluster of straw hits associated to a particle trajectory is identified using a robust helix fit algorithm to perform the pattern recognition. A Kalman based track fitter
provides a precise geometric and kinematic estimate 
of the electron track parameters, resulting in a momentum resolution of about 200 keV/$c$ for CE. 

\subsection{The Electromagnetic Calorimeter}

\begin{figure}[htbp]
\centering
\includegraphics[width=0.9\textwidth]{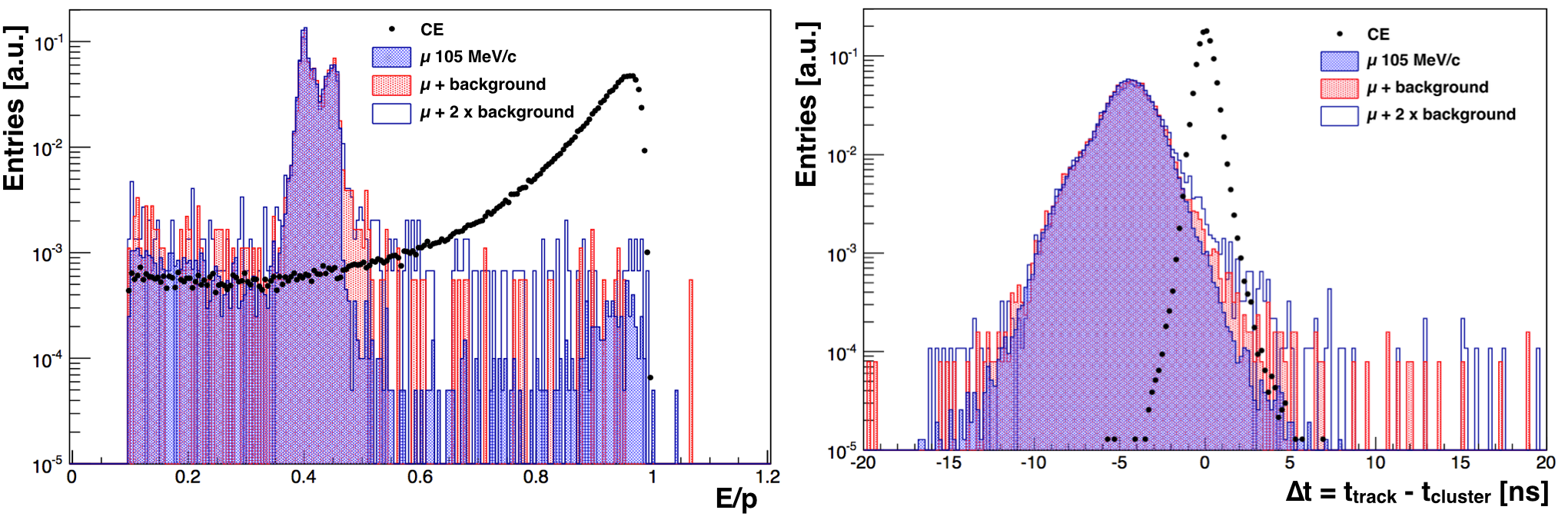}    
\caption{\scriptsize{Distribution in $\Delta$t = t$_{track}$ - t$_{cluster}$, on the left, and E/p, on the right, for cosmic electrons and 105 MeV/$c$ muons reconstructed as electrons.
Different colors correspond to different assumption of the background level. \cite{Pezzullo:2016tym}} \label{img:calopid}}
\end{figure}

\noindent The crystal calorimeter consists of about 1400 undoped 3.4$\times$3.4$\times$20 cm$^{3}$ CsI crystals, arranged in two disks. The disks are separated by 75~cm, half of the 
conversion electron $\lambda$, in order to maximize the calorimeter acceptance. Each crystal is coupled with two 2$\times$3-arrays of 6$\times$6~mm UV-extended SiPM. 
This configuration ensures at 100 MeV an energy resolution better than the 
10$\%$ and a time resolution better than 500~ps. The calorimeter complements the tracker providing: a powerful particle identification, 
a seed for the pattern recognition in the tracker and an independent software trigger system. The particle identification provides a good separation between CEs and muons un-vetoed by the CRV 
and mimicking the signal. The required 
muon rejection factor is provided with 95$\%$ efficiency on the signal, combining the time of flight difference between the tracker track and the calorimeter cluster with the E/p 
ratio, see Figure \ref{img:calopid}, into a simple likelihood. 

\section{Expected Background}

\begin{figure}[htbp]
\centering
\includegraphics[width=0.8\textwidth]{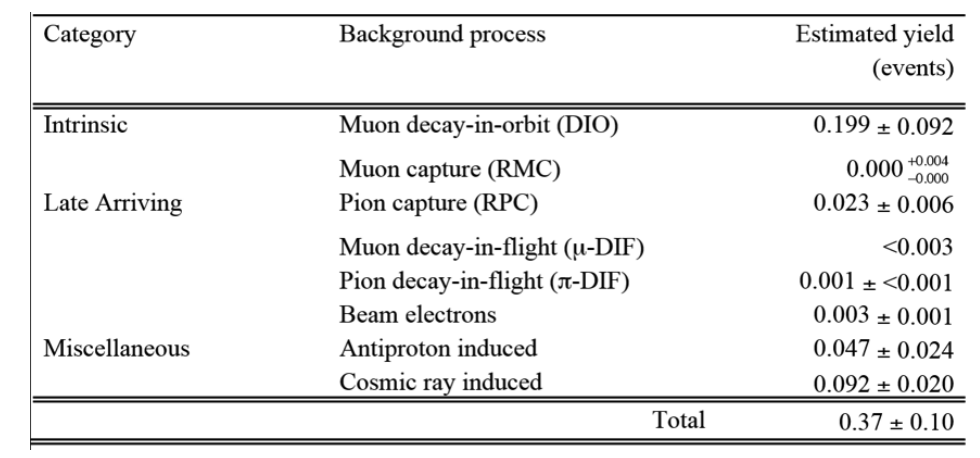}    
\caption{\scriptsize{Backgrounds yield expected by Mu2e in three years of data taking, corresponding to 6.8$\times$10$^{17}$ stopped muons. \cite{MU2ETDR}} \label{img:backtable}}
\end{figure}

\noindent Muons that stop in the stopping target can undergo beta decay in the field of the nucleus (decay in orbit, DIO), be captured (ordinary muon capture, OMC) or convert 
to electrons. For Aluminium, the muon DIO probability is about 39$\%$. The kinematic limit for the muon decay in vacuum is at about 54 MeV, but the recoil of the nucleus generates 
a long tail that has the endpoint exactly at the conversion electron energy. This tail falls rapidly down as $(E_{CE} - E)^{5}$ and only 10$^{-17}$ of the DIO energy spectrum is within 1 MeV 
on the left of the endpoint. This represents the main background source for Mu2e and scales linearly with the beam intensity. So the only technique to suppress DIO is to carefully 
measure the momentum with the high resolution tracker.  The OMC, which has a probability of 61$\%$, is not a source of background but for a very small contribution 
from photons produced in radiative muon captures (RMC), as reported in Table \ref{img:backtable}. Backgrounds arising from interactions at the production target are overwhelmingly prompt
and arrive at the stopping target in time with the muon beam. As shown in Figure \ref{img:beamstruct}, these backgrounds are eliminated by defining a delayed selection windows. However,
out-of-time protons impinging on the production target can produce background that come anyway in the search window. A proton extinction factor of 10$^{10}$ is needed to keep these Late 
Arrival Backgrounds (LAB) at a neglegible level. In Mu2e it is obtained including in the proton extraction line a custom AC dipole that sweeps out-of-time protons into a collimator. 
The remaining background sources consists of cosmic rays and antiprotons. Cosmic rays can interact in the DS producing delta rays or can be trapped directly 
in the DS magetic field and leave a track in the tracker equal to a conversion electron helix. The cosmic ray veto system ensures that the first type of cosmic rays 
background is kept under control, while the second component is suppressed by means of the particle identification capabilities of the detector system. Antiprotons from interaction in the PS can drift slowly
until they annihilate. These annihilations produce high multiplicity final states that can generate background in time with the search window. To limit antiprotons induced
background, two Be absorbers are placed at the entrance and in the middle of the TS. Table \ref{img:backtable} summarizes the expected backgrounds yield for three years of data taking \cite{MU2ETDR}, corresponding to 6.8$\times$10$^{17}$ stopped muons. The total 
estimated background yeld is 0.4 events and the expected sensitivity is $R_{\mu e}$ < 6$\times$10$^{-17}$ at 90$\%$ of C.L. .    

\section{Status}

\noindent The Mu2e experiment has recently received the CD-3 approval from the U.S. Department of Energy. The R$\&$D program is complete and all the detectors are frozen as a 
final design. The first large scale prototypes are under construction and beam tests are scheduled for the fall 2016. The installation phase will begin in 2017, while data taking 
is expected to start in early 2021.

\section{Conclusions}

\noindent Mu2e will search for the CLFV process $\mu^{-}$ Al $\rightarrow$ e$^{-}$ Al with an unprecedented sensitivity, improving by a factor $10000$ the current best limit. If 
a signal is observed, that will be undeniable proof of new physics and it will provide data complementar to LHC and to the other CLFV experiments. If no signal is found, the expected upper 
sensitivity $R_{\mu e}$ < 6$\times$10$^{-17}$ will set constrains to many models at mass scale up to thousands of TeV. 

\section*{Acknowledgments}

\noindent We are grateful for the vital contributions of the Fermilab staff and the technical staff of the participating institutions. This work was supported by the US Department of 
Energy; the Italian Istituto Nazionale di Fisica Nucleare; the US National Science Foundation; the Ministry of Education and Science of the Russian Federation; the Thousand Talents 
Plan of China; the Helmholtz Association of Germany; and the EU Horizon 2020 Research and Innovation Program under the Marie Sklodowska-Curie Grant Agreement No.690385. Fermilab is 
operated by Fermi Research Alliance, LLC under Contract No.\ De-AC02-07CH11359 with the US Department of Energy.

\end{document}